\documentclass[%
 aip,
 amsmath,amssymb,
reprint,%
]{revtex4-1}

\usepackage{graphicx}
\usepackage{dcolumn}
\usepackage{bm}
\usepackage{float}
\usepackage[utf8]{inputenc}
\usepackage[T1]{fontenc}
\usepackage{mathptmx}
\newcommand{\BS}{Bi$_2$Se$_3$}
\newcommand{\BSS}{Bi$_2$Se$_3$ }

\begin{document}

\preprint{AIP/123-QED}

\title{Study Of Surface Spin-Polarized Electron Accumulation In Topological Insulators Using Scanning Tunneling Microscopy}
\author{S. Tyagi}
\affiliation{Department of Electrical and Computer Engineering, University of Maryland, College Park, MD 20742, USA}
\author{M. Dreyer}
\affiliation{Department of Physics, University of Maryland, College Park, MD 20742, USA}
\author{D. Bowen}
\author{D.Hinkel}
\affiliation{Laboratory for Physical Sciences, College Park, MD 20740, USA}
\author{P.J. Taylor}
\affiliation{Army Research Laboratory, Adelphi, MD 20783, USA}
\author{\\ A. L. Friedman}
\affiliation{Laboratory for Physical Sciences, College Park, MD 20740, USA}
\author{R.E. Butera}
\author{ C. Krafft}
\affiliation{Laboratory for Physical Sciences, College Park, MD 20740, USA}
\author{I. Mayergoyz}
\affiliation{Department of Electrical and Computer Engineering, University of Maryland, College Park, MD 20742, USA}

\date{}

\begin{abstract}
The results of scanning tunneling microscopy experiments using iron-coated tungsten tips and current-carrying bismuth selenide (\BS) samples are reported. Asymmetry in tunneling currents with respect to the change in the direction of bias currents through \BSS samples has been observed. It is argued that this asymmetry is the manifestation of surface spin-polarized electron accumulation caused by the ninety-degree electron spin-momentum locking in the topologically protected surface current mode. It is demonstrated that the manifestation of surface spin-polarized electron accumulation is enhanced by tin doping of \BSS samples. Furthermore, the appearance of spin-dependent density of states in current carrying \BSS samples has also been observed. 
\end{abstract}

\maketitle
\section{Introduction}

Topological insulators are currently a very active area of research in physics in general, and spintronics in particular due to their unique physical properties \cite{ando,collTI,moore2010birth,qiandzhang7} and promising engineering applications  \cite{STT2, app3,QD}. Topological insulators have a bulk band gap like an ordinary insulator (or semiconductor) and conducting surface states topologically protected by time-reversal symmetry. These surface conducting states exhibit ninety-degree locking\cite{collTI} between the electron spin and its momentum caused by very strong spin-orbit interaction in these materials. This ninety-degree locking results in surface accumulation of spin-polarized electrons when a bias current flows through the topological insulator.

In this paper, we report the experimental study of these surface spin-polarized electron accumulations by using scanning tunneling microscopy (STM) with iron-coated tungsten tips. It was observed that there is a change (i.e., asymmetry) in the tunneling current with respect to the change in the direction of the bias current through the topological insulator. It can be reasoned that this asymmetry is caused by the following two factors. The first is the change in the spin orientation of surface electrons caused by the change in bias current direction. The second is the spin-dependent density of states of the iron-coated tungsten tip. Thus, it can be concluded that the above asymmetry reveals local surface accumulation of spin-polarized electrons caused by the ninety-degree spin-momentum locking. 

Our experiments were performed using MBE-grown \BSS samples. These are binary compounds which represent the second-generation of topological insulator materials \cite{collTI} with a relatively large bulk band gap of around 0.3eV and the simplest (almost ideal) surface band structure with a single Dirac cone for the (topologically protected) conducting surface mode. In \BSS samples, the unintentional bulk conductivity results in bulk currents which obscure the surface conducting states\cite{hsieh2009tunable, dopSn}. It has been shown that chemical doping of intrinsic topological insulators with such elements as calcium \cite{hsieh2009tunable,dopCa} or tin \cite{dopSn,dopeSn2} moves the bulk Fermi level into the bulk band gap. This results in reduced conduction due to bulk states. In our experiments, \BSS samples with different levels of tin (Sn) doping were used to reduce bulk conductivity. It was observed that the increase in tin doping levels results in the increase of tunneling current asymmetry with respect to the direction change of the bias current through the samples. The latter indicates the enhanced manifestation of the topological surface mode. It was also found that the current flow through the \BSS  samples results in the appearance of spin-dependent density of states in the samples. This was revealed by spin-polarized electron tunneling from the iron-coated tungsten tips to the samples. The appearance of this spin-dependent density of states in samples may be the result of inclined (i.e., not horizontal) crossing of the Dirac cone by the surface Fermi level in the presence of the bias current through \BSS samples.    
 
The STM technique has been extensively used for the study of physical properties of topological insulators \cite{STM1,STM2,STM3, STM4,STM5,STM6}. This is because it is very local (nanoscale) in nature, which is its clear advantage in comparison with other electrical and optical measurements\cite{collTI, moore2010birth}. The contributions of this paper are the local STM measurements of surface spin-polarized electron accumulations in \BSS samples with various tin doping levels by using iron-coated tips.

\section{Technical Discussion}

\begin{figure}[h]
 	\includegraphics[scale=0.3]{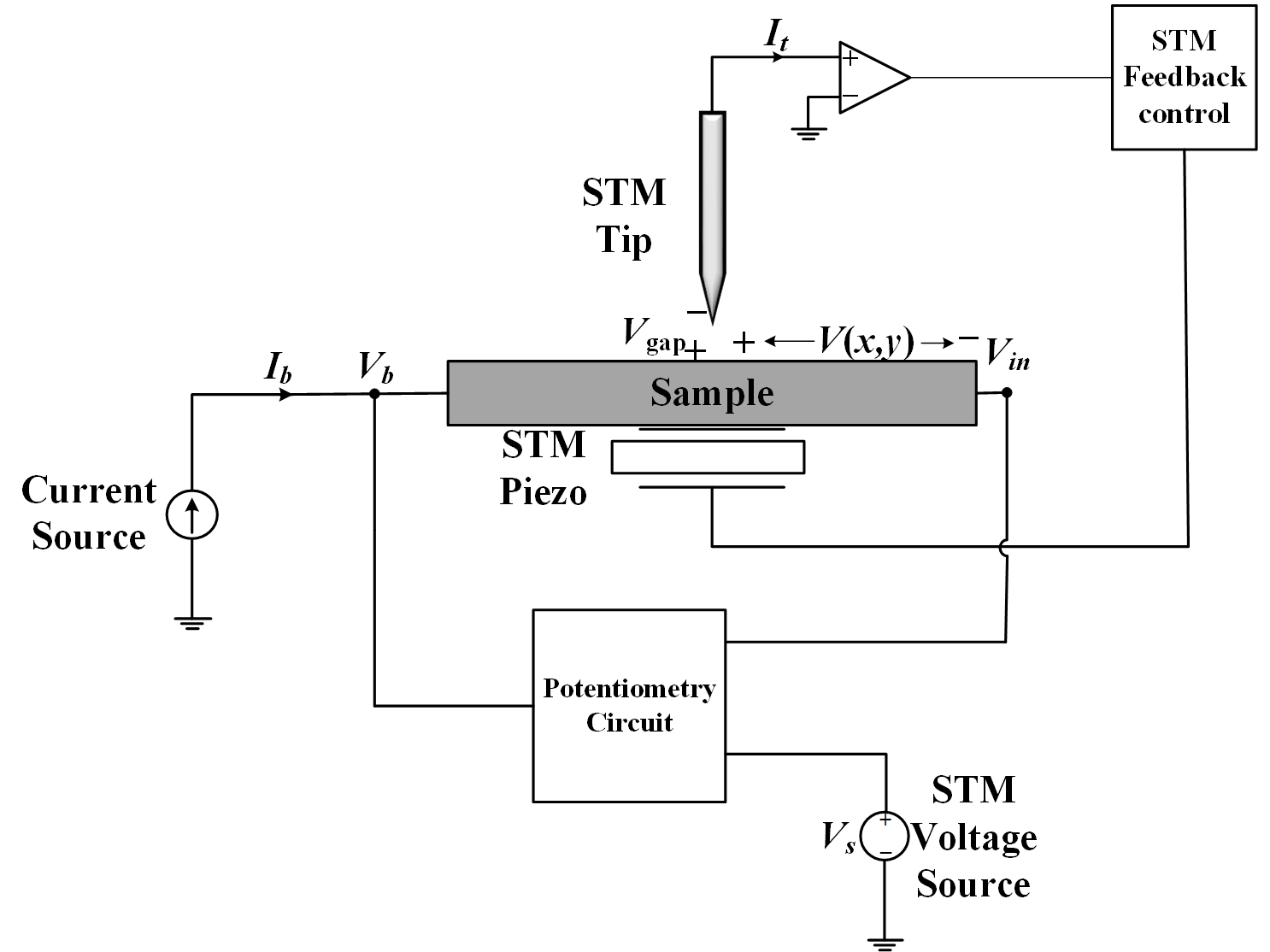}
 	\caption{Schematic of STM set-up.}
 	\label{sche}
 \end{figure}

Our STM studies were performed in a two chamber Omicron ultra high vacuum (UHV) STM system, using the experimental setup schematically shown in Fig.~\ref{sche}. The current source shown in this figure was used to provide the desired bias current through the \BSS sample, while the voltage source was used to apply desired tunneling voltage between the STM tip and the sample. The bias current results in a voltage drop along the sample, and this complicates the application of the proper tunneling voltage ($V_{\text{gap}}$) between the tip and the sample. This problem was solved by developing the special potentiometry technique \cite{xie1}, and thereafter it was used in the STM study of the spin Hall effect \cite{xie2,xie3,xieaip}. Potentiometry ensures that $V_{\text{gap}} = V_{s}$ even in the presence of bias-currents through the sample.
 
The iron-coated tungsten tips were fabricated in UHV by using e-beam deposition of iron on clean tungsten tips. Energy dispersive X-ray spectroscopy (EDS) of these tips confirmed the presence of iron at their front ends.

The \BSS thin-films with the thickness of about 400 nm were grown by MBE on GaAs substrates. These films were cut to 0.5 cm in width and 1 cm in length.  Along with intrinsic \BSS samples, three tin-doped samples were used with different levels of tin dopings. These three Sn-doped samples have their stoichiometry of the form (Bi$_{1-x}$Sn$_{x}$)$_2$Se$_3$ with $ x$ being equal to $0.0073, 0.0216$ and $0.049$. Below, these samples shall be referred to as samples A, B and C, respectively. To achieve clean surfaces of these \BSS samples for scanning, they were cleaved in vacuum and then placed in the UHV STM chamber. 

\begin{figure}[h]
	\includegraphics[width=\columnwidth, height=0.18
 \textheight]{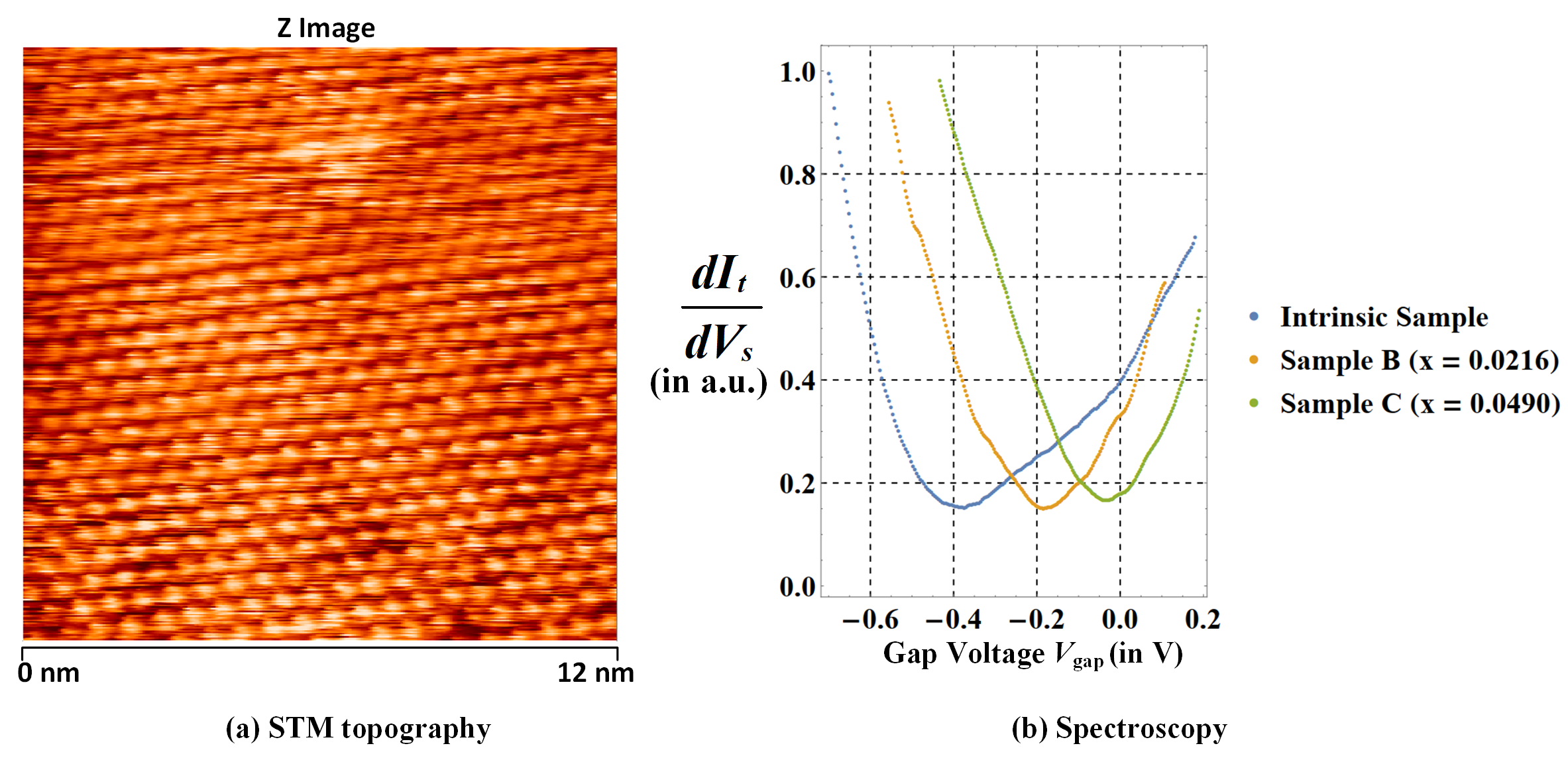}
	\caption{(a) Atomic resolution image for intrinsic \BSS sample at the tunneling current of 300 pA and bias voltage 200 mV. This is the image of Se atoms on the surface of \BS. Average distance between atoms was measured to be 4.07 \AA, consistent with values reported in literature\cite{STM6}.  (b) STM spectroscopy Dirac cone measurements for different levels of tin dopings. These curves were obtained using the standard lock-in measurement technique.}
	\label{aanddidv}
\end{figure}

Before conducting the experiments with current carrying samples, STM images of the atomic structure of the \BSS sample surfaces were obtained along with spectroscopy measurements of the Dirac cone of the topologically protected surface conducting mode. Typical surface morphology for intrinsic (undoped) \BSS samples is shown in Fig.~\ref{aanddidv}(a). Typical spectroscopy curves for \BSS samples are shown in Fig.~\ref{aanddidv}(b), and are consistent with measurements previously reported\cite{STM1,STM4}. Note here that the Dirac point corresponds to the minimum of the dI/dV curve. Measurements shown in Fig.~\ref{aanddidv}(b) reveal the shifting of the Dirac point towards zero tunneling voltage with the increase in the doping levels.  

Next, the STM experiments with current carrying \BSS samples were performed. As the initial step, a DC bias current through the sample in one direction was introduced, and the constant tunneling current mode of STM tip was used to determine the achievement of thermal equilibrium. Then, the bias voltage ($V_{\text{gap}}$) and tunneling current ($I_t^{\leftarrow}$) were adjusted to desired values. Subsequently, the constant tunneling current feedback for STM tip was turned off, the direction of the bias current was abruptly changed to the opposite direction, and the tunneling current ($I_t^{\rightarrow}$) was measured as a function of time. The superscript arrows in the above notations refer to the direction of the bias current through the samples.

First, the above STM experiments were performed with tungsten (not iron-coated) tips with a bias current of 20 mA, and no significant changes in tunneling currents were observed upon changing the direction of bias currents through the \BSS samples (i.e., $I_t^{\rightarrow}/I_t^{\leftarrow} \approx 1$). This implies that in the case of tungsten tips, tunneling currents are even-symmetric functions of bias currents.

\begin{figure}[h]
	\includegraphics[width=\columnwidth, keepaspectratio]{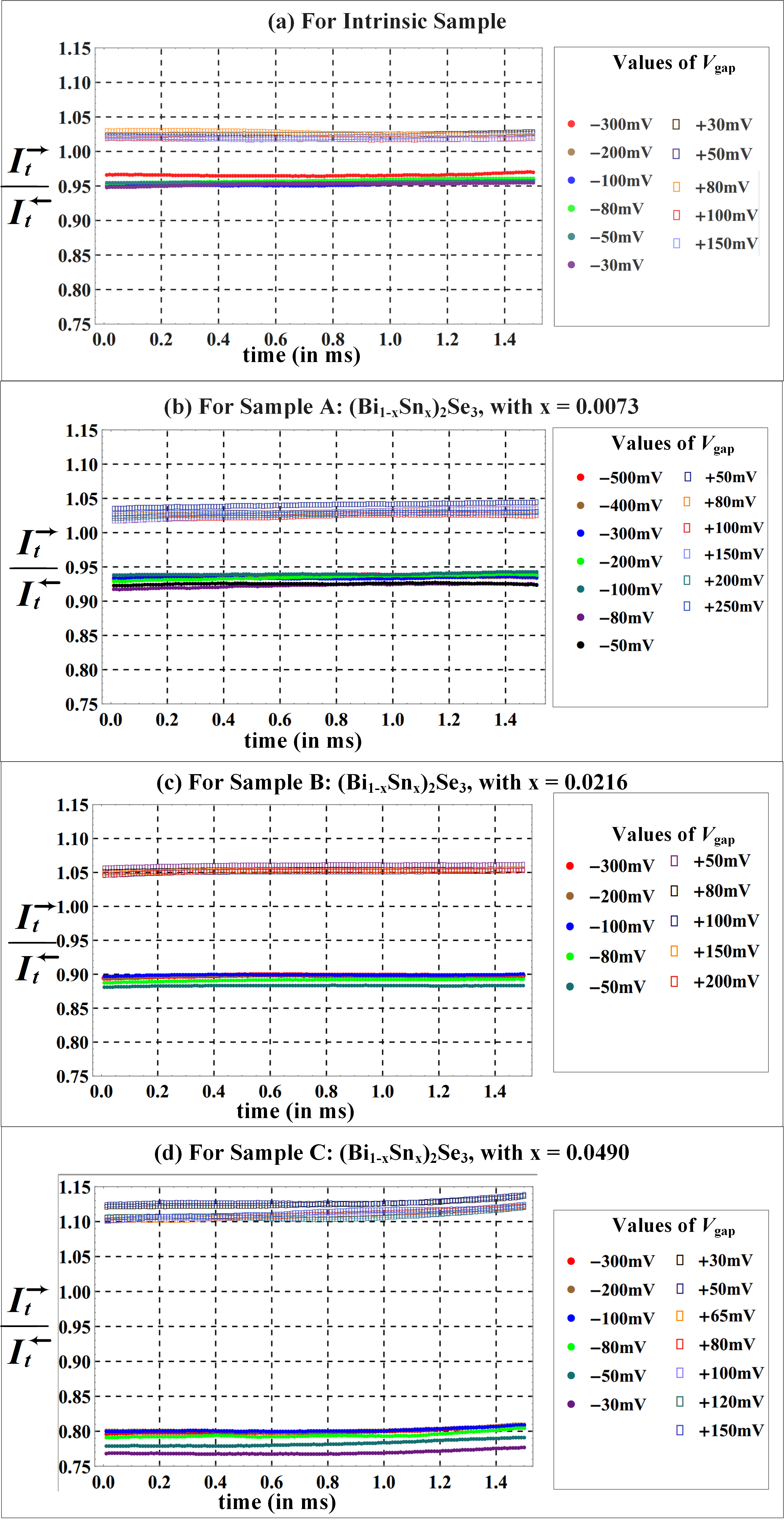}
	\caption{Observed asymmetry in tunneling curent for differently doped samples and $I_t^{\leftarrow} = 100$ pA. Steady state (in time) measurements of tunneling currents are presented.}
	\label{mainres}
\end{figure}

Then, the STM experiments were conducted by using iron coated tungsten tips. They were performed for a bias current of 20 mA through the \BSS samples and tunneling currents ($I_t^{\leftarrow}$) of about 100 pA. It was found that for iron-coated tips, the tunneling current is not an even-symmetric function of the bias current. The asymmetry is characterized by the ratio $I_t^{\rightarrow}/I_t^{\leftarrow}$ of the tunneling currents measured for the opposite direction of the bias current.  The experimentally observed asymmetry of the tunneling currents for different values of tunneling voltages is presented in Fig.~\ref{mainres}(a) for the intrinsic \BSS sample and in Figs.~\ref{mainres}(b), \ref{mainres}(c) and \ref{mainres}(d) for tin-doped \BSS samples A, B and C, respectively. 

It is important to note that, since no asymmetry $I_t^{\rightarrow}/I_t^{\leftarrow}$ in tunneling currents was observed in the case of tungsten tips, the reported asymmetry in Fig.~\ref{mainres} can be attributed to the spin-dependent tunneling of electrons between the Fe-coated tip and current-carrying \BSS sample. It is also evident that the asymmetry appreciably depends on the polarity of the tunneling voltage. 

\begin{figure}[h]
	\includegraphics[width=\columnwidth, keepaspectratio]{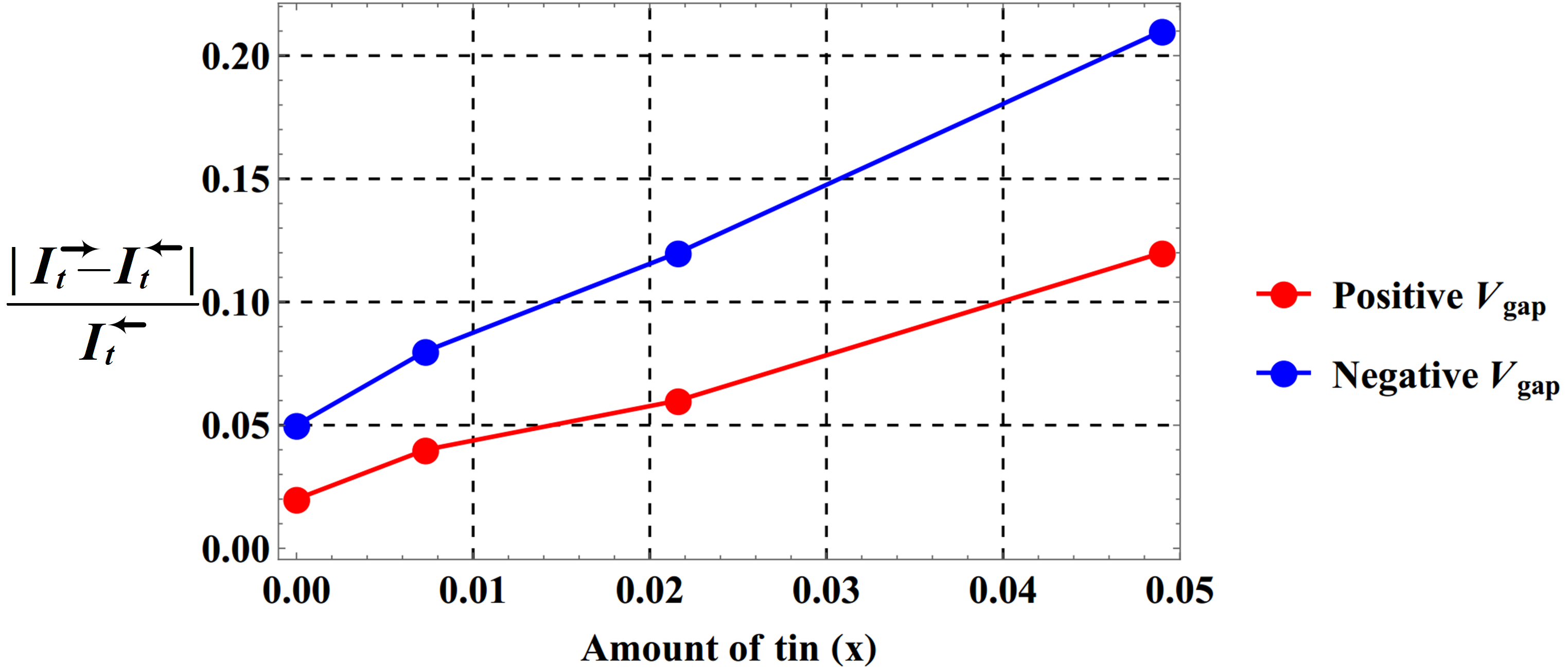}
	\caption{Asymmetry as a function of tin doping measured at $I_t^{\leftarrow} = 100$ pA and $V_{\text{gap}} = \pm 50$ mV.} 
	\label{inc}
\end{figure}

In Figs.~\ref{mainres}(a)-(d), the bottom curves (with $I_t^{\rightarrow}/I_t^{\leftarrow} < 1$) correspond to the polarity of tunneling voltages for which electrons tunnel from the current carrying \BSS samples to the iron-coated tip. On the other hand, the top curves (with $I_t^{\rightarrow}/I_t^{\leftarrow} > 1$) correspond to the opposite polarity of tunneling voltages for which spin-polarized electrons tunnel from the iron-coated tip to the current carrying sample. The bottom curves reveal the surface spin-polarized electron accumulations caused by the ninety-degree locking between momentum and spin of electrons of the topologically protected surface conducting mode in \BSS samples. Indeed, this locking results in the reversal of electron spin orientation upon reversal of the bias current direction. This spin reversal leads to the change in tunneling currents due to the spin-dependent density of states of the iron-coated tip. 

The top curves in Figs.~\ref{mainres}(a)-(d), corresponding to the spin-polarized electrons tunneling from the iron-coated tip, reveal the spin-dependent density of states in current carrying \BSS samples. The appearance of this spin-dependent density of states may be the result of inclined (not horizontal) crossings of the Dirac cone by surface Fermi levels\cite{siu2018spin,hus2017detection}. These inclined crossings are caused by bias surface currents in \BSS samples. These bias surface currents are increased with the increase in tin doping levels of the sample. The latter results in larger slope angles of surface Fermi levels, which leads to the increase in the spin dependent density of states in current carrying \BSS samples. This is evident from the top curves in Figs.~\ref{mainres}(a)-(d). \par

Furthermore, it is clear from the Fig.~\ref{inc} that the asymmetry in tunneling currents is increased with the increase in tin doping levels, from about 5\% for the intrinsic sample to 21\% for sample C (for the case of negative $V_{\text{gap}}$). This can be explained as follows. Tin doping increases the bulk resistance, but does not affect the physical properties of the conducting surface mode because it is topologically protected. This leads to the increase of the surface portions of bias currents through \BSS samples with higher doping and, consequently,  results in the observed increase in tunneling current asymmetry.\par

\begin{figure}[h]
	\includegraphics[width=0.7\columnwidth]{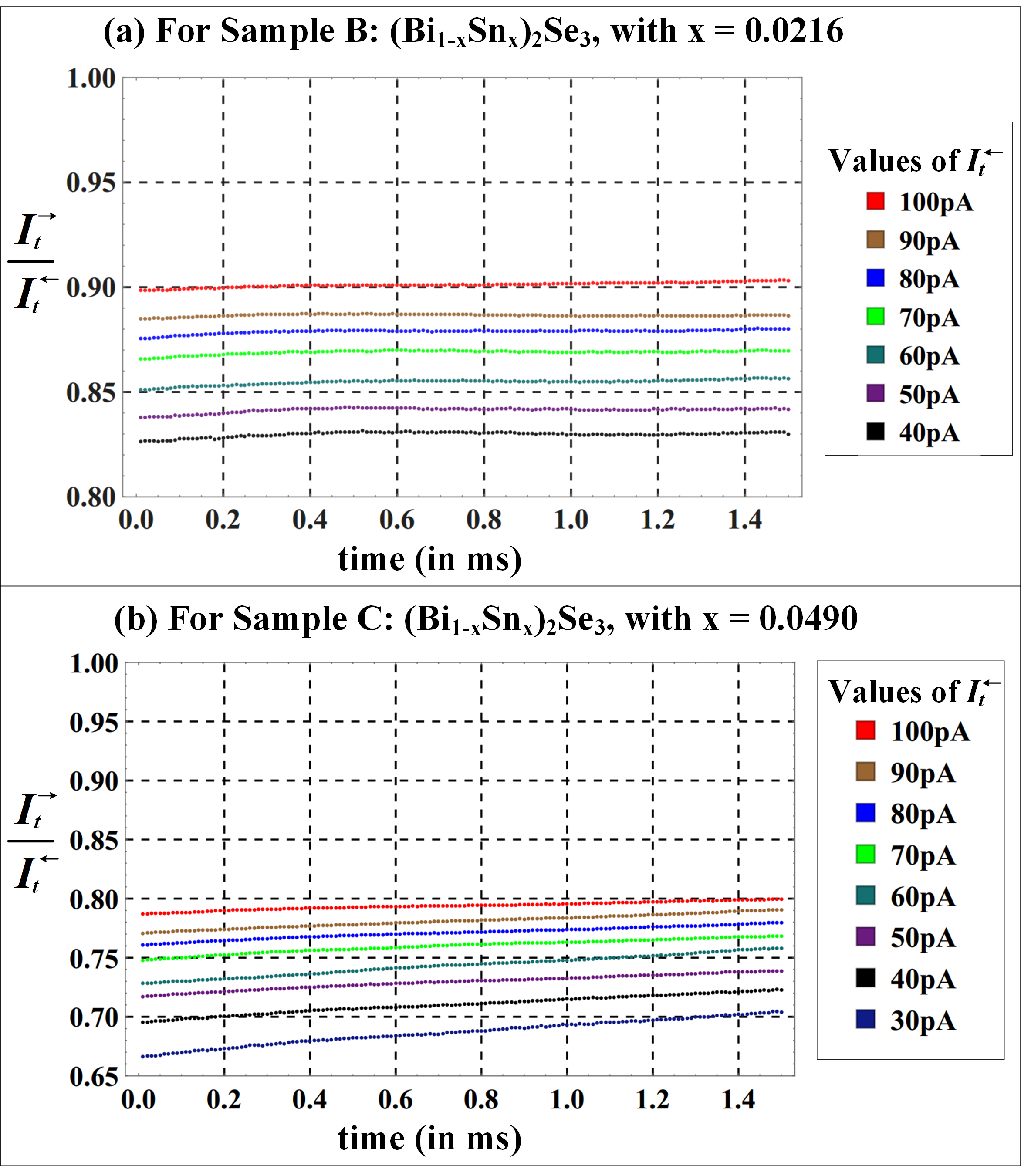}
	\caption{Asymmetry for different values of $I_t^{\leftarrow}$ and $V_{\text{gap}} = -100$ mV}
	\label{itinc}
\end{figure}

It has also been observed that the described asymmetry in tunneling currents is enhanced for smaller values of these currents. This is illustrated by Fig.~\ref{itinc}(a) and \ref{itinc}(b) for samples B and C, respectively. It is evident from this Fig.~\ref{itinc}(b) that the above asymmetry may reach about 30\% for the tunneling current of around 30 pA in the case of sample C. This increase in the observed asymmetry suggests that spin-polarized electron accumulation becomes more pronounced with a reduction in tunneling current $I_t^{\leftarrow}$. This may suggest that the effect of density of states on the value of tunneling currents becomes more dominant with the decrease of these currents.

\section{Conclusion}

In this paper, a novel experimental technique for the STM detection of spin-polarized electron accumulation on the surface of topological insulators is presented. The observed changes (asymmetries) in the values of the tunneling currents upon reversal of the direction of bias currents through \BSS samples are clear experimental signatures of surface spin-polarized electron accumulations. Furthermore, the observed increase in the aforementioned asymmetry with the increase in tin doping levels of \BSS samples reveals the suppression of bulk conductivity and consequent enhancement of surface spin-polarized electron accumulations in \BS.

The presented results suggest that the STM based measurements with iron-coated tungsten tips may open new opportunities in the study of surface effects in topological insulators. These measurements are very local in nature (i.e. they are on the nanoscale). Hence, the described technique can be extended to study the correlation of these effects with surface morphology of topological insulators. 

\bibliography{References}
\end{document}